\documentclass[epj]{svjour}
\newif\ifpdf            
\ifx\pdfoutput\undefined \pdffalse \else \pdftrue \fi
\ifpdf 
\usepackage[pdftex]{graphicx}
\pdfcompresslevel9 \DeclareGraphicsExtensions{.jpg,.pdf,.png,.mps}
\else 
\usepackage[dvips]{graphicx}
\DeclareGraphicsExtensions{.eps,.ps} \fi
\usepackage{bm,amssymb}

\newcommand{\beq}{\begin{equation}}
\newcommand{\eeq}{\end{equation}}

\begin{document}

\title{Phase chirality and stereo-selective swelling of cholesteric elastomers}
\titlerunning{Stereo-selectivity of cholesteric elastomers}
\author{S.~Courty, A.R.~Tajbakhsh and E.M.~Terentjev}
\authorrunning{Courty,  Tajbakhsh and Terentjev}
\institute{Cavendish Laboratory, University of Cambridge,
Madingley Road, Cambridge CB3 0HE, U.K. }

\date{\today}

\abstract{ Cholesteric elastomers possess a macroscopic ``phase
chirality'' as the director $\bm{n}$ rotates in a helical fashion
along an optical axis $z$ and can be described by a chiral order
parameter $\alpha$. This parameter can be tuned by changing the
helix pitch $p$ and/or the elastic properties of the network. The
cholesterics also possess a local nematic order, changing with
temperature or during solvent swelling. In this paper, by
measuring the power of optical rotation $d \Psi/d z$, we discover
how these two parameters vary as functions of temperature or
solvent adsorbed by the network. The main result is a finding of
pronounced stereo-selectivity of cholesteric elastomers,
demonstrating itself in the retention of the ``correct'' chirality
component of a racemic solvent. It has been possible to quantify
the amount of such stereo-separation, as the basic dynamics of the
effect.
 \PACS{
  {33.55.Ad }{Optical activity, optical rotation; circular dichroism} \and
  {61.30.Vx}{Polymer liquid crystals} \and
  {87.80.Pa}{Morphometry and stereology}}
} 

\maketitle

\section{Introduction}
The nature appears to be inherently chiral. From the atomic scale
with asymmetric carbon, to much larger length scales -- like our
hands or even spiral galaxies, all have the same common feature of
lacking the inversion symmetry, while not characterised by any
vector (dipolar) property. In other words, many natural objects
are non-superimposable with their mirror image and define a pair
of opposite handedness,  right and left. This is the notion of
chirality. Since its first discovery in 1848 by Pasteur
\cite{pasteur} and attempts on mathematical abstraction by Kelvin
\cite{kelvin}, chirality and more particularly molecular chirality
have always been a source of interest in various fields, from
mathematics to medicine. Even though it is now much better
understood, this breaking of symmetry is still an active and
exciting field of research across disciplines. It is important to
realise that handedness is not an absolute concept; its
quantitative characteristics depend on the property being observed
\cite{Osipov95,Harris99}, the origin of many questions and
disagreements between different groups of results.

A word on terminology is due here. There are several ways of
describing chiral substances, developed in chemistry. The Rosanoff
(1906) notation distinguishes between D[+] (for {\it dextra}) and
L[--] (for {\it laevo}) on the basis of relative arrangement four
different bonds of the chiral carbon. The Cahn-Ingold-Prelog
(1956) notation is also based on ranking of bonds according to
specific sequence rules, improved such that it can be used for
more complex molecules; it specifies R[+] (for {\it rectus},
clockwise rotation) and S[--] (for {\it sinister}, anticlockwise).
Sometimes different sources of organic chemistry data have mixed
notations, e.g. the Aldrich catalogue quotes cholesterol
derivatives as R-(--), with specific optical rotation
$[\alpha]_D=-40^{\rm o}$. This simply reflects the fact that
different chirality indices (scalar and tensorial) may be
introduced to describe different physical responses, while the
proper notation is not yet developed in spite of many recent
advances. So, the sense of steric chirality (asymmetry in the
geometric shape of the object) is not necessarily the same as that
of the third-order dielectric polarisability $\beta_{ijk}$
(determined by electronic structure) and that, in turn, may be
different at different frequencies. As a result, the chiral
intermolecular interaction may not be of the same handedness as,
e.g., the rotation of light polarisation. The macroscopic ``phase
chirality'' of cholesteric structure studied here is a result of
cooperative action of all such effects and we simply distinguish
it by macroscopic optical rotation, right-handed (clockwise, R) or
left-handed (anticlockwise, L).

Chirality has a dramatic impact on most aspects of life, as
enantiomers often have differences or even opposite properties
(e.g. different odours, different biological functionality,
toxicity, etc). This is a real problem for pharmaceutical
industry, or foods and cosmetics, since the synthesis of a
chemical compound leads to a racemic mixture (equal proportions of
right- and left-handed enantiomers) and the stereo-selection is
always a very difficult task. The reason for the difficulty lies
in the fact that a pair of enantiomers differs in shape and
electronic properties only in a very subtle way, which results in
small corrections only to high-order molecular polarisability. As
a result the molecular interactions that are sensitive to the
handedness are always very weak. New methods have been developed
recently to measure such forces, for example, by detecting a
difference in adhesion between an AFM tip coated with chiral
molecules and the left- or right-handed substrate
\cite{McKendry98}.

One of the main techniques in the field of chiral
stereo-separation is column chromatography, in which a racemic
mixture diffuses at slightly different rates through a silica gel
coated with a molecular layer of specific chirality. Recently, a
new concept of stereo-selection was introduced, based on the
macroscopic phase chirality in topologically imprinted cholesteric
networks \cite{Mao:01b,Courty:03}. If one were to quantify the
phase chirality, a corresponding order parameter has to be
introduced that would reflect this symmetry. One traditional
example is the cholesteric helix in liquid crystalline systems,
where the molecular units are locally aligned, on average, in a
uniaxial (nematic) fashion, but with the director coherently and
periodically rotating over a larger length scale (helical twist),
Fig.~\ref{helix}. In analysing the imprinted cholesteric
elastomers, a system that may retain the phase chirality while not
having any on the molecular level \cite{Mao:00}, Mao and Warner
(MW) have introduced a parameter that measures the (inverse)
strength of imprinted helicity in the polymer network, $\alpha =
\sqrt{K_2/D_1}q_0$, where $K_2$ is the Frank (twist) elastic
constant \cite{deGennes93}, $q_0$ is the helix wavenumber at
network formation and $D_1$ is the relative-rotation coupling
constant \cite{WT03}. If the network is formed with a large
$\alpha$, it would not be able to sustain its helical twisting
when the chiral molecular moieties are removed, while at $\alpha
\ll 1$ the more rigid elastic network retains most of the
imprinted helix.

There are several ways of monitoring the state of phase chirality
in a cholesteric material. A traditional method based on selective
reflection at a certain wavelength of light is not chirality
specific. In fact, recent studies has shown that one can generate
the bandgap for both right- and left- circular polarisations of
incoming light by only a slight mechanical deformation of
cholesteric elastomers \cite{Bermel:02,Cicuta:02}. Instead we
concentrate on the effect of optical rotation. The classical
optical activity (Faraday effect) of a solution of chiral
molecules can be enhanced by up to $10^4$ times in a cholesteric
phase generated by the same amount of chiral dopant $\Phi$
\cite{deGennes93,deVries:51}. As the helical pitch $p = 2\pi/q$ is
inversely proportional to the concentration of chiral dopant
$\Phi$ (see section~\ref{theo} below), the cholesteric phase would
be very sensitive to any small variation of this concentration in
the network, resulting in uniform twist-untwist changes.

In this paper, we will show how to utilise this macroscopic
enhancement of normally weak chiral interactions to generate
stereo-selectivity of cholesteric elastomers with respect to
different components of a racemic solvent. In contrast to the
earlier work on imprinted networks, here we study the naturally
cholesteric elastomers, where the chiral molecular moieties are a
part of the network. Therefore, we cannot separate the effects of
molecular chiral interactions, and those due to the phase
chirality, so cleanly as in imprinted systems. However, since the
results are broadly similar to those reported in \cite{Courty:03},
we believe the macroscopic coherence of order parameter
modulations in the helix plays a dominant role. In any case, for
practical purposes of developing an efficient system for
stereo-selective separation of racemic solvents, the cholesteric
elastomers could be a preferred option due to the ease of their
preparation.

One central issue we shall be struggling with throughout this work
is obvious, but has not been systematically examined in this
context before. As one adds a solvent to a liquid crystalline
network, whether a racemic mixture or a general achiral solvent,
the magnitude of the local nematic order parameter changes
(usually, decreases). This results in a rapid change in local
optical birefringence (affecting the optical rotation) and also
the strength of phase chirality (reducing the specific interaction
with chiral solvent). As soon as the material becomes isotropic,
i.e. loses its coherent cholesteric structure altogether, it also
loses the stereo-selectivity (at least to the accuracy of our
detection methods). This, in a way, is a proof that the phase
chirality determines the stereo-selective swelling, and not the
specific interaction with molecular moieties (which are still
there in the isotropic phase). In much of the section~\ref{res} we
shall be challenged by the competition between this local liquid
crystalline order and the macroscopic phase chirality, aiming to
develop a set of analytical tools to quantify the results.

\section{Theoretical background} \label{theo}
\begin{figure}
\resizebox{0.42\textwidth}{!}{\includegraphics{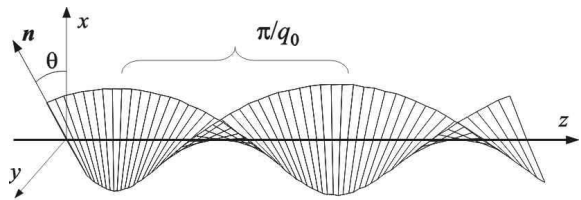}}
\caption{Spatial distribution of the director $\bm{n}$ in an ideal
cholesteric helix along the optical axis $z$. Because of the local
quadrupolar symmetry of the nematic order, the periodicity
interval is only the half-pitch, $\pi/q_0$.} \label{helix}
\end{figure}
Locally cholesteric is an amorphous uniaxially ordered medium,
like a nematic liquid crystal, described by the local order
parameter $Q_{ij}= Q (n_{i}n_{j}-\frac{1}{3}\delta_{ij})$.
However, on larger scales the director $\bm{n}$ is a periodic
modulated function of coordinates, in the ideal state rotating
along a single axis $z$: $n_{x}=\cos\theta, n_{y}=\sin\theta,
n_{z}=0$. With it rotates the local dielectric ellipsoid, with
principal refractive indices $m_{\rm o}$ along the director and
$m_{\rm e}$ perpendicular to $\bm{n}$ (the dielectric anisotropy
$\Delta m=m_{\rm o} - m_{\rm e}$ is directly proportional to the
magnitude of order parameter $Q$). In the classical cholesteric
helix the azimuthal angle is $\theta=q_{0}z$, with the
corresponding pitch $p_{0}= 2\pi/q_{0}$, see Fig.~\ref{helix}. The
breaking of inversion symmetry results from the presence of chiral
molecular groups in the material. If the elastomer network is
crosslinked in this state, freezing in the helical pitch $p_0$,
any further change in the concentration of chiral groups in the
network would give rise to the elastic free energy of the form:
\begin{equation}
F_{\rm el} = \int {\textstyle{\frac{1}{2}}} \left[ K_{2}
({\textstyle{\frac{d}{dz}}}\theta - q)^{2} +D_{1}\sin^{2} (\theta-
q_{0}z) \right] dz ,   \label{f1}
\end{equation}
per unit area in the $x$-$y$ plane. Both terms represent the
penalty for deviating from the initial state with $\theta=q_0z$.
The rubber-elastic contribution simply records the conformation at
network crosslinking as its ground state. The Frank term has a
minimum determined by the current preferred state of phase
chirality: the average helix wavenumber $q=2\pi/\langle p\rangle =
4\pi \beta \Phi $, where $\Phi$ is the total concentration of
chiral molecular groups in the material (assumed small to maintain
the linear relationship $q \propto \Phi$) and the coefficient
$\beta$ is the measure of microscopic twisting power of these
groups \cite{deGennes93}.

Since in this paper we are mostly concerned with the effects of
swelling by solvents added to the crosslinked elastomer network,
two more physical effects have to be taken into account. The
swollen network resists to stretching of its chains; if the
overall volume is increased from $V_0$ to $V=V_0(1+\phi)$ due to
an extra concentration of added solvent $\phi$, the effective
dilatation strain $\gamma \approx (1-\phi)^{-1/3}$ contributes to
the rubber elasticity. When a small concentration $\phi$ of
``impurity'' is added to the mesogenic system, the phase
transition temperature shifts down in a linear fashion, $T^*
\approx T_c (1- \kappa \, \phi)$. As a result, the additional free
energy arises (per unit area):
\begin{eqnarray}
F_{\rm s} &=& \int {\textstyle{\frac{1}{2}}} \left[ 3 \mu
[(1-\phi)^{-2/3}-1] + A_{\rm o}
[T-T^*(\phi)] \, Q^2  \right] dz \nonumber \\
 &\approx & \int {\textstyle{\frac{1}{2}}} \left[  2 \mu \, \phi +
A_{\rm o} [T-T_c(1-\kappa \, \phi)] \, Q^2  \right] dz \, .
\label{f2}
\end{eqnarray}
Here $\mu$ is the rubber modulus. The ($\phi$-dependent)
$Q^2$-term shows the leading contribution to the thermodynamic
Landau-de~Gennes expansion in powers of the local order parameter.
One also has to consider the mixing entropy and the coupling due
to the Flory $\chi$ interaction parameters and thus complete the
analysis of swelling by solvents. In fact, the problem is much
more delicate. In an anisotropic material one cannot assume the
simple volume change -- instead the principal directions (along
$z$ and in the $x$-$y$ plane) would stretch by slightly different
factors, dependent on the nematic order through chain anisotropy
\cite{Wang:97}. Also, MW show \cite{Mao:01b} that the coupling
constant $D_1$ is renormalised on swelling. However, the values of
swelling and uniaxial strain in our experiments are so small that
the corrections are only minor, reaching 2\% at the most;
accordingly we neglect this additional anisotropy and many
complexities and subtle physical effects associated with it.

When the solvent added to the cholesteric elastomer is achiral
(not optically active) the only effect on the pitch is through the
affine expansion of sample dimension along $z$. In taking this
view we assume that the average pitch does not change with the
reduction of order parameter $Q(\phi)$. One may not regard it
obvious, or even correct, because there is a large literature on
cholesteric liquid crystals showing the variation of pitch with,
e.g., temperature. Our assumption is supported, at least in our
elastomers, by the direct observation of nearly constant pitch,
quoted in Fig.~\ref{Dn-psi=f(temp)} below, see also the literature
data \cite{Maxein:99}.

When we work with a racemic mixture of right- and left-handed
enantiomers and generate an imbalance between the components,
$\Delta\phi = \phi_L-\phi_R$ (with the total $\phi = \phi_L +
\phi_R$), the additional molecular chirality modifies the
cholesteric pitch as well:
\begin{equation}
\langle {p}\rangle \equiv \frac{2\pi}{\langle q \rangle} = 2\pi
\left[ (1-\phi)^{1/3}q_0 - q_{\rm s} \Delta \phi \right]^{-1} ,
\label{p}
\end{equation}
where $q_{\rm s}$ is a coefficient measuring the interaction of
specific chirality of the solvent with the network. Note that the
affine expansion term in (\ref{p}) assumes that all three
dimensions of the network are swelling in the same proportion,
$\gamma=(1-\phi)^{-1/3}$. In our experiments, the samples will
conserve their area in the $x$-$y$ plane and would only change
thickness along $z$, by $\gamma_z=(1+\phi)$, thus modifying
equation~(\ref{p}).

Our purpose in this paper is to explore the stereo-selectivity of
a cholesteric elastomer, leading to the imbalance $\Delta\phi$ of
enantiomers swelling the network, by independently monitoring the
weight and shape of the sample (providing the data on total
$\phi$) and the changes in optical rotation (giving direct access
to $\Delta \phi$). In order to interpret the results we need to
revise the classical results of de Vries \cite{deVries:51} on the
rotating power of cholesteric helix.

By solving the eigenproblem for electromagnetic waves propagating
along $z$ one finds the superpositions of circularly polarised
plane waves of opposite signs. Their phase difference gives the
optical rotation $\Psi$ in the medium, or its rotating power per
unit length along $z$. The problem is, in fact, much more delicate
than it is frequently presented in the literature, because of the
need to correctly treat the boundary conditions for the incident
linearly polarised light, see e.g. \cite{Belyakov79}. Another
important issue, not treated well in the original de Vries'
approach, is the limit of vanishing phase chirality, $\langle q
\rangle \rightarrow 0$. This corresponds to the so-called Mauguin
limit, or the `waveguide regime', when $ \langle p \rangle \Delta
m \gg \Lambda_0/\bar{m}$, with $\Lambda_0$ is the wavelength of
incident light and $\bar{m} = \sqrt{ \bar{\varepsilon}} =
\sqrt{}\frac{1}{2}(\varepsilon_\| + \varepsilon_\bot)$ the average
refractive index of the material. Not going into great detail of
these complicated problems with over 30 years of history, for our
practical purpose of analysing the rotation of plane polarisation
in a cholesteric elastomer we shall use a simplified result for
the rotation rate (`rotatory power') $d \Psi /dz$. Fortunately, in
our system, the parameters combine in such a way that we never
cross the bandgap. Accordingly, the difference between the full de
Vries solution,
 \begin{eqnarray}
\frac{d\Psi}{d z} &=& \frac{2\pi}{\langle p \rangle} \left( 1+
\frac{1}{2\lambda'} \left[\sqrt{1+
\lambda'^2-\sqrt{\delta^2+4\lambda'^2}}  \right.
\right. \label{DeVries} \\
&& \qquad \qquad \left. \left. -\sqrt{1+
\lambda'^2+\sqrt{\delta^2+4\lambda'^2}} \right]\right) \ ,
\nonumber
 \end{eqnarray}
and its more familiar approximate expansion
\cite{deGennes93,deVries:51} is minor. In Eq.~(\ref{DeVries}) the
combination $\lambda'=\Lambda_0/\langle p \rangle \bar{m}$ is the
non-dimensional ratio of the light wavelength to the pitch, and
$\delta =(\varepsilon_{\bot}-\varepsilon_{\|})/
(\varepsilon_{\bot} +\varepsilon_{\|}) \approx \Delta m/\bar{m}$
is the parameter of relative dielectric anisotropy. In our
material the extraordinary and ordinary refractive indices are,
respectively, $m_{\rm e}=1.75$ and $m_{\rm o}=1.6$ (making
$\bar{m}=1.68$ and the light wavelength in the medium
$\Lambda_0/\bar{m} = 377\, $nm). For $\lambda' = \sqrt{1 \pm
\delta}$ a dispersion anomaly appears in (\ref{DeVries}) in the
form of Bragg-like reflection; this full solution does not have a
divergence for $d\Psi/dz$ but a finite value on the edges of a
bandgap of width $\delta$ \cite{deVries:51}. In our case this
anomaly would center at $\langle p\rangle \approx 377\,
\hbox{nm}$. The analysis below will show that the initial pitch of
the cholesteric helix is $ p_{0} \approx 580 \, \hbox{nm}$, which
means that our range of measurements is always on the longer
wavelength side of the bandgap.

\begin{figure}
\resizebox{0.45\textwidth}{!}{\includegraphics{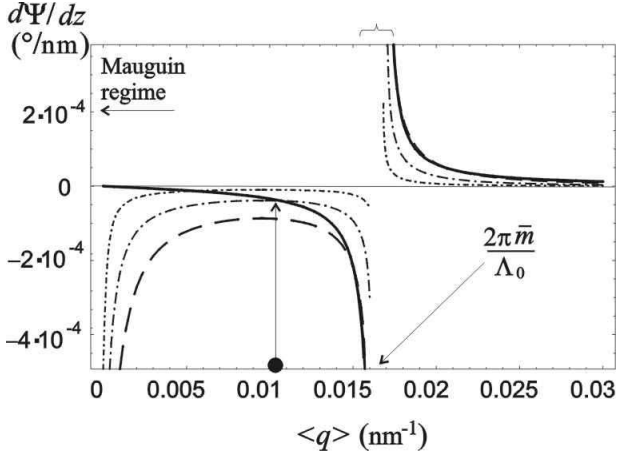}}
 \caption{The
rate of optical rotation $d\Psi/d z$, as function of the helix
wavenumber $\langle q \rangle =2\pi \bar{m} \lambda'/\Lambda_0$.
The solid line shows the interpolated result with correct limiting
behaviour. The broken lines show the classical de Vries plots for
decreasing local birefringence ($\Delta m =0.15$, $\Delta m =0.05$
and $\Delta m =0.01$). The bandgap is at $\lambda' =1$, with a
width decreasing with $\Delta m$. The dot marks the initial
cholesteric pitch $p_0=580$\,nm.} \label{deVries plot}
\end{figure}

Figure~\ref{deVries plot} shows the variation of the rotatory
power with the helix wavenumber (or equivalently, with $\lambda'$
for fixed $\Lambda_0$) for the parameters of our experimental
system. We plot the function that interpolates between the correct
behaviour near the bandgap and the required linear decrease at
$q\rightarrow 0$. For comparison, a sequence of de Vries curves is
also plotted, for a series of decreasing $\Delta m$ (and the
underlying nematic order parameter $Q$, directly proportional to
it).

Equation~(\ref{DeVries}) and all its modifications contain two
important parameters which are changing in our experiments: the
relative dielectric anisotropy $\Delta m$ and the effective pitch
length $\langle p \rangle$. The first of these changes with
temperature and swelling with small-molecule `impurities' and
reflects the local (nematic) liquid crystalline ordering. The
changing effective pitch, or the helical wavenumber $\langle q
\rangle = 2\pi/\langle p \rangle$, reflects the phase chirality of
a cholesteric and is determined by chiral imbalance between the
components of racemic mixture swelling the cholesteric network.
Rotating power is an unambiguous measure of phase chirality (as
opposed to, say, selective reflection spectrum which can be too
broad in cholesteric networks and also present in both right- and
left-handed modes for the same material \cite{Cicuta:02}). We
shall use Eq.~(\ref{DeVries}) to extract the effective pitch from
the measurements of optical rotation. For this, let us re-write it
explicitly showing the two relevant parameters, $\Delta m$ and
$\langle p \rangle$, for $\delta \ll 1$:
\begin{equation}
\frac{d \Psi}{d z} \approx -\frac{ \pi \bar{m}^2 \Delta
m^{2}\langle p \rangle^3} {2 \Lambda_0^2 \left( \bar{m}^2 \langle
p\rangle^2- \Lambda_0^2 \right) } \ . \label{dpsidz}
\end{equation}
To find $\langle p \rangle$ we need to resolve this cubic
equation, which gives the approximate result in the relevant
region of parameters, represented by the solid line in
Fig.~\ref{deVries plot}:
\begin{equation}
\langle p \rangle \approx -\frac{\pi \Delta m^2 + \sqrt{\pi^2
\Delta m^4+ 16 \bar{m}^2 \Lambda_0^2 (d\Psi/dz)^2}}{4 \bar{m}^2
(d\Psi/dz)} \label{pitch=f(Dn,psi)}
\end{equation}
This interpolated model will serve us for the rest of this work,
to help extracting the values of effective cholesteric pitch, as a
measure of phase chirality, from the measured $d\Psi/dz$ and the
deduced $\Delta m$.

\section{Methods}

\subsection{Sample preparation}
\begin{figure} [b]
\resizebox{0.42\textwidth}{!}{\includegraphics{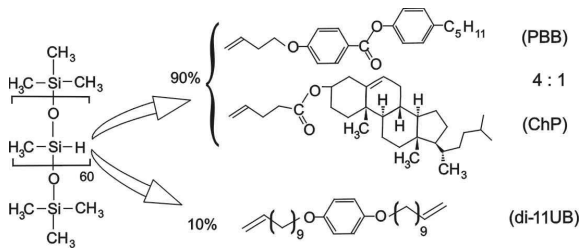}}
\caption{Chemical composition of the cholesteric network.}
\label{chemistry}
\end{figure}
The cholesteric liquid crystal elastomer was synthesised following
the general method introduced by Kim and Finkelmann \cite{Kim:01}.
Siloxane backbone chains were reacted under centrifugation at
7000rpm with 90\,mol\% mesogenic side groups (the mixture of
nematic 4-pentylphenyl-4'-(4-buteneoxy)benzoate, labelled PBB, and
cholesterol pentenoate, ChP, in proportion 4:1) and 10\,mol\% of
1,4\,di(11-undeceneoxy)benzene, di-11UB, crosslinker groups for 45
minutes at 75$^\circ$C to form a partially crosslinked gel. For
the further 4 hours the reaction proceeded under centrifugation at
60$^\circ$C, during which time the solvent was allowed to
evaporate, leading to an anisotropic deswelling of the gel and
completion of crosslinking. All of the volume change in this setup
occurs by reducing the thickness of the gel, while keeping the
lateral dimensions fixed (due to centrifugation): this introduces
a very strong effective biaxial extension in the $x$-$y$ plane. At
this second-stage temperature of 60$^\circ$C the dried polymer is
in the cholesteric phase and its director is forced to remain in
the plane of stretching -- this results in the uniform cholesteric
texture which is finally crosslinked at this second stage of
preparation. The chemical composition of our networks is shown in
Fig.~\ref{chemistry}.

Differential scanning calorimetry measurements (Perkin-Elmer Pyris
7 DSC) were used to characterise the resulting elastomer. The
glass transition was unambiguously determined at $T_g \approx
-10^\circ$C and the clearing point, the isotropic-cholesteric
transition occurs at $T_{\rm c}\approx 90^\circ$C. No additional
thermal transitions were found between these two critical
temperatures. All experiments were performed at room temperature,
sufficiently far from both transitions (except when we studied the
variation of optical rotation with order parameter, for
comparison).

\subsection{Experimental set-up}
\begin{figure} 
\resizebox{0.42\textwidth}{!}{\includegraphics{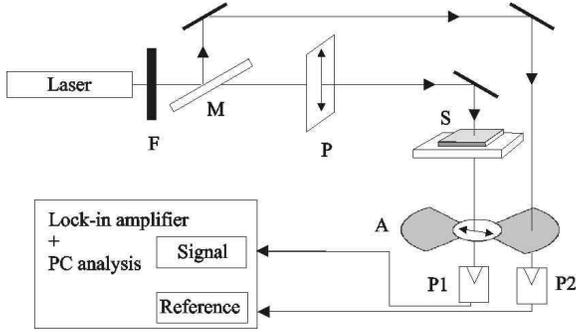}}
\caption{Experimental set-up for measuring the optical rotation
$\Psi$. (F): neutral density filter; (M): beam splitter, (P):
Glan-Thomson polariser; (S): glass cover slip with sample; (A):
rotating analyzer with light chopper; (P1,P2): photodiodes linked
to the lock-in amplifier.} \label{apparatus}
\end{figure}
The optical rotation $\Psi$ can be determined experimentally by
using a dynamical method \cite{rotref} based on measuring the
phase difference between the split parts of linearly polarised
laser beam (He-Ne laser, $\lambda_{laser}=633 nm$,~$30mW$, from
Melles-Griot), one passing through the sample and the rotating
analyzer (fixed frequency $\sim 16 Hz$), the other through the
optical chopper (providing the reference signal to lock on),
Fig.~\ref{apparatus}. The phase difference $\Delta \Theta$ between
the two beams is measured by an integer number of periods with a
lock-in amplifier (Stanford Research) and corresponds directly to
the optical rotation $\Psi$ from which the effective cholesteric
pitch $\langle p \rangle$ is then calculated. The elastomer is
deposited onto a glass coverslip (area conservation will be
observed) and a solvent droplet of known volume ($10 \mu l$) is
placed on it, with the beam spot in the middle. The solvents used
for achiral and racemic environment are respectively
toluene:hexane mixture (ratio 1:6, from Acros) and 2-Bromopentane
(from Acros).

The optical configuration described above has been also used to
measure the birefringence $\Delta m$ of a very similar nematic
elastomer during solvent evaporation. From this we obtained an
independent data on  $\Delta m$ to use in the analysis based on
Eq.~(\ref{pitch=f(Dn,psi)}), assuming that the local properties of
cholesteric are approximately the same as of the corresponding
nematic. In the measurements of optical rotation we do not care
about the ellipticity of light out of the sample, only about the
angle of the principal axis. In the $\Delta m$ measurements of a
uniformly birefringent medium we do. The optimal configuration is
to send linearly polarised light at an angle of $\pi/4$ to the
director and then place a $\lambda/4$ plate after the sample with
one of its axis adjusted to be parallel to the incident
polarisation (to recover the linear polarisation) \cite{Lim:78}.
The relative phase difference $\Delta \Theta$ between the
perpendicular and parallel polarisations is directly related to
$\Delta m$ by $\Delta \Theta = 2\pi d\Delta m/\Lambda_0$, with $d$
the independently measured sample thickness.

For temperature measurements, the glass coverslip is directly
placed onto a hot stage connected to a controller (Stanton
Redcroft) with a range of temperature varying from $20^{\rm o}C$
to $200^{\rm o}C$.

\section{Results and discussion} \label{res}

\subsection{Optical rotation}
\begin{figure} [b]
\resizebox{0.42\textwidth}{!}{\includegraphics{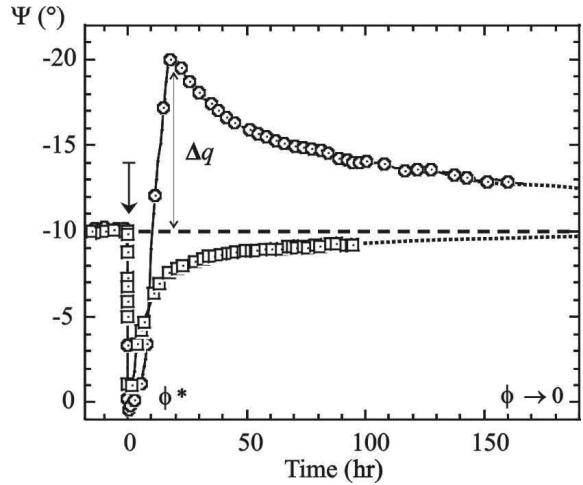}}
\caption{Variation of total angle  $\Psi$ of rotation of incident
linear polarisation after swelling of cholesteric elastomer with
achiral (squares) and racemic solvent (circles). The arrow
represents the time deposition of the solvent droplet onto the
sample.} \label{Dpsi}
\end{figure}
Figure~\ref{Dpsi} shows the evolution with time of the optical
rotation from cholesteric elastomer after solvent swelling and
subsequent drying. Initially the value of $\Psi$ is constant and
is determined by the initial pitch $p_{0}$ of the dry cholesteric
state and the local nematic order defined by $\Delta m_{0}$. At
$t=0$ a drop of solvent is placed in the beam path, as labelled by
arrow in the plot. The optical rotation rapidly drops to zero
because the amount of solvent taken into the network exceeds the
concentration $\phi^*$ at which the material becomes isotropic,
$Q(T,\phi)=0$. Accordingly, the rotatory power vanishes -- see
Eq.~(\ref{dpsidz}) with $\Delta m = 0$. The spectacular difference
between the two types of solvent is observed on subsequent slow
evaporation. As the concentration of the achiral solvent in the
swollen network decreases, the liquid crystalline order returns
back (cf. Fig.~\ref{Dn} below for detail) and the observed optical
rotation gradually approaches the same level it had in the initial
dry cholesteric elastomer. As the total concentration $\phi$ of
the racemic solvent decreases, the rotatory power returns, but its
magnitude is significantly higher than in the dry cholesteric.
After a maximum at $\sim 20$~hr, even after a very long time, the
network does not return to its original helical state, but
saturates asymptotically at $\Psi \approx -12^{\circ}$. This rise
of optical rotation above its initial value, and the whole
non-monotonic time (and indirectly -- concentration) dependence,
are the signature of chiral separation of solvent components. The
network retains the component that matches its own phase
chirality, while letting the other component evaporate -- as a
result the effective helical power increases. This is labelled by
$\Delta q$ in the plot, to make connection with Eq.~(\ref{p}) and
Fig.~\ref{p=f(phi))} below.

The behavior presented on figure.~\ref{Dpsi} is remarkable and
represents the central result of this work. Our challenge is now
to quantitatively analyse the effect of stereo-selectivity of
cholesteric elastomers. We need to extract from the optical data,
$d \Psi/d z$, the information on how the pitch $\langle p \rangle$
varies as function of the solvent $\phi$ retained by the network.
For this, we need to know the relationship between the effective
pitch $\langle p \rangle$ and the local order $Q$ (or $\Delta m$)
and as the network approaches the transition into an isotropic
state, first by heating above $T_c$ and secondly, by swelling
above $\phi^*$. This will lead to a way to determine the current
cholesteric pitch any small variation of solvent in the network
and to relate it to chiral imbalance $\Delta \phi$ of guest
molecules in the network. Finally, we can monitor the overall
amount of solvent in the network by simply measuring the weight of
samples as function of time. This will provide the independent
data on $\phi(t) = \phi_L + \phi_R$. In addition, this is
important to provide the information about the current thickness
of the swollen network, $d(t)$, which is necessary to convert the
raw rotation angle $\Psi$ into the rotatory power $d\Psi/dz$;
since our samples have their $x$-$y$ area conserved, the thickness
change is simply $d(t)=d_0 (1+\phi)$.

\subsection{Local order parameter}

It is unfortunately nearly impossible to independently measure the
local birefringence $\Delta m$, or equivalently, the local nematic
order parameter $Q$, of a cholesteric liquid crystal. The
difficulty is the same as to measure $Q$ in a polydomain texture.
Perhaps the only experimental technique that could offer access to
such local information is NMR, relating the angularly averaged
line splitting to local bias of probe molecule orientation,
affected by the nematic mean field. In our case, all we can do is
to measure $\Delta m$ on a chemically similar nematic liquid
crystal system and assume that its value and variation would be
the same in a cholesteric. It is not a totally unreasonable
assumption: the degree of nematic order is very reliably $Q \sim
0.5 \pm 0.1$ for most nematic liquid crystal materials (apart from
main-chain polymers, which is not our case). The refractive
indices depend more strongly on the molecular structure, varying
between, say, $1.45$ and $1.85$ in different nematic materials.

We chose a composition of nematic elastomer as close to that in
Fig.~\ref{chemistry} as we could: with PBB and the same amount of
di-11UB crosslinker, with hopefully a similar molecular anisotropy
and the amount of polarisable $\pi$-electrons. As a confirmation
of our choice, the clearing temperature of this material,  $T_c
\approx 90^{\rm o}$C, was similar to that of the cholesteric.
Making an aligned monodomain nematic elastomer with this chemical
composition, we then measure its birefringence as function of
temperature. Since the anisotropy of dielectric tensor is directly
related to the nematic order parameter, $\Delta m = {\rm
const}\cdot  Q$ \cite{deGennes93}, we obtain the latter by
calibrating the proportionality constant at room temperature
against a separate X-ray measurement of $Q$.

\begin{figure}
\resizebox{0.42\textwidth}{!}{\includegraphics{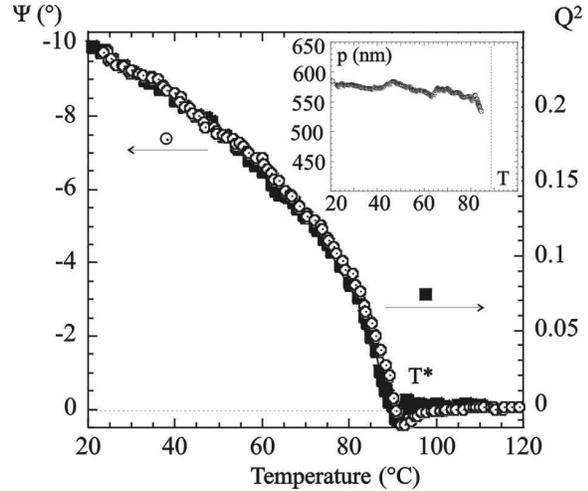}}
\caption{Superposition of $\Psi$ (circles) and $Q^{2}$ (dark
squares) as function of the temperature, from room temperature to
$T^{*}$. In inset, dependence of the pitch $p$ with the
temperature, proving that the nematic order parameter has no
strong effect on the average cholesteric pitch, at least in our
elastomers. } \label{Dn-psi=f(temp)}
\end{figure}
Figure~\ref{Dn-psi=f(temp)} gives the results and the
corresponding analysis. We plot, against temperature, the square
of the nematic order parameter, $Q^2$, determined as described
above (filled squares, right $y$-axis). This shows the expected
behaviour, reported and discussed many times in the recent
literature \cite{WT03}. The data for $Q(T)$ can be very well
fitted by an empirical critical dependence $Q=0.8(1-T/T_c)^{0.28}$
(an interesting and provocative observation of its own, also
discussed in the literature). This data set is compared with the
result plotted on the left $y$-axis (open circles), for the
optical rotation of our cholesteric elastomer as it was heated
towards its isotropic phase (note, the initial rotation of $\sim
10^{\rm o}$ is the same as that in Fig.~\ref{Dpsi}). The matching
is quite spectacular, including even the pretransitional region
where both $\Psi$ and $\Delta m$ effectively change sign, proving
that $\Psi \propto Q^2$, which is the result predicted by
Eq.~(\ref{dpsidz}).

This exact matching of optical rotation $\Psi (T)$ and the local
birefringence $\Delta m (T)$ also proves that the other parameters
in the Eq.~(\ref{dpsidz}), in particular the cholesteric pitch
$\langle p \rangle$, do not significantly change with the nematic
order. The inset in Fig.~\ref{Dn-psi=f(temp)} shows the effective
pitch calculated from the data for $\Psi$ and $\Delta m$ by our
method outlined in section~\ref{theo}. It clearly does not vary
very much, in spite of the underlying nematic order continuously
dropping to zero. Such a conclusion is not unexpected in
cholesteric elastomers where the helix is crosslinked into the
rubbery network, and has been reported before \cite{Maxein:99}.

\begin{figure}
\resizebox{0.42\textwidth}{!}{\includegraphics{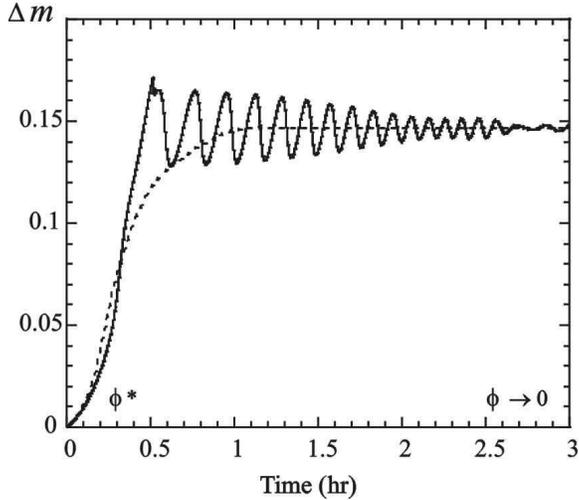}}
\caption{Variation of the birefringence $\Delta m$ on solvent
evaporation from the nematic elastomer, between the isotropic
phase (at $\phi^{*}, \ t=0$) to the nematic phase ($\phi
\rightarrow 0$). The dotted line represents a smoothed model curve
for $\Delta m(t)$.} \label{Dn}
\end{figure}

Now, what will happen if the cholesteric network is swollen with a
non-chiral small molecule solvent, at constant temperature? First
of all, as for the temperature, we need to know what is the
dependence of the underlying local order parameter $Q$ on the
solvent concentration $\phi$. It was assumed that, for low $\phi$,
the effect of added impurities is to linearly shift down the
transition temperature $T_c=T^*(\phi)$, see Eq.~(\ref{f2}). It is
easy to find the critical solvent concentration at which the
material becomes isotropic, $\phi^* \approx 8\%$ in our case,
giving the parameter $\kappa \approx 2.3$. Accordingly, we might
have expected that the order parameter follows the law $Q(\phi) =
0.8 (1-T/T_c[1-\kappa \phi])^{0.28}$, with the variation exactly
as that shown in Fig.~\ref{Dn-psi=f(temp)}. Instead, we directly
measured $\Delta m$ as function of time as the droplet of solvent
deposited on the nematic sample (see the configuration
Fig.~\ref{apparatus}) swells the network and then gradually
evaporates. Figure~\ref{Dn} presents the result, in terms of
$\Delta m$ rather than $Q$, with the time $t=0$ chosen at the
moment when the sample first returns to its birefringent state (at
$\phi^*\approx 8\%$, measured independently). As the solvent
evaporates further ($\phi \rightarrow 0$), the birefringence
increases and eventually returns to its initial value
demonstrating the complete restoration of the nematic order.

The actual experimental result shown in Fig.~\ref{Dn} is very
interesting and rather unexpected, and will certainly lead to a
separate investigation. We observe periodic oscillation of $\Delta
m$ as $\phi \rightarrow 0$ in a nematic elastomer deswelling,
while constrained on a substrate. We hypothesise that these
oscillations are caused by the concentration gradient as one side
of the sample is attached to a glass substrate (and its area
conserved); as the solvent evaporates from the free surface, it
also causes a gradient of $Q$, and with it -- the mechanical
strains in the network, which attempts to uniaxially expand along
$\bm{n}$ with the strain magnitude $\propto Q$. Clearly, the set
of coupled differential equations for the time evolution of local
$\phi$, $Q$ and elastic strain has an oscillating instability.
Note that the cholesteric network, in our main experiments on
swelling and optical rotation is also attached to a glass
substrate, so presumably a similar effect should occur there.

Although tempted, here we do not discuss this phenomenon further,
not to distract the reader from the main theme of this paper. For
our analysis of effective helical pitch/wavenumber as a measure of
phase chirality, we need to substitute the raw data for $d\Psi
/dz$ obtained from Fig.~\ref{Dpsi} and the data for $\Delta m$
from Fig.~\ref{Dn} into the Eq.~(\ref{pitch=f(Dn,psi)}). If one
compares the time scales, it becomes clear that the curious
oscillating regime of $\Delta m(t)$ occurs at a very early stage
of deswelling, in a region where $\Psi \approx 0$ in
Fig.~\ref{Dpsi}. Accordingly, for the main (and most interesting)
bulk of data we could use a simpler, smoothed model for $\Delta
m(t)$ which is shown as a dotted line in Fig.~\ref{Dn}. We
obtained such a model in a somewhat arbitrary fashion, trying to
match as closely as possible the data and the linear model for
$Q(T,\phi)$; however, the main point of this argument is that in
the region of main interest the deviation of the model from the
raw data is not relevant.

\subsection{Quantifying the stereo-selectivity}

\begin{figure}
\resizebox{0.42\textwidth}{!}{\includegraphics{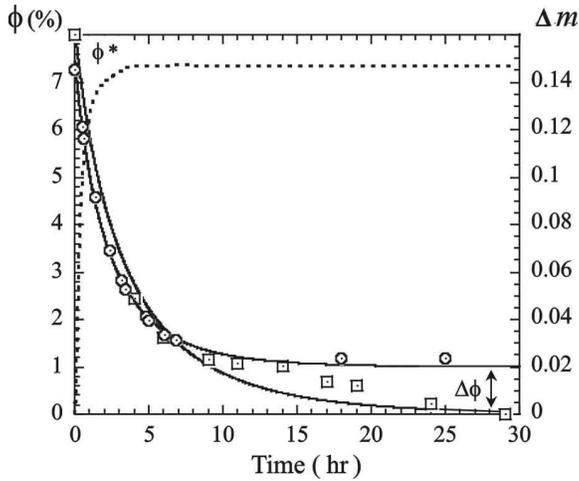}}
\caption{Time evolution of total concentration $\phi(t)$ as the
solvent evaporates from the cholesteric network. As in
Fig.~\ref{Dpsi}, squares represent the achiral solvent and circles
-- the racemic mixture. The time $t=0$ is chosen, as in
Fig.~\ref{Dn}, as a moment when the material first re-enters the
liquid crystalline state (at $\phi^* \approx 8\%$. Solid lines are
the fit by a stretched-exponential model (see text). The right
$y$-axis shows the parallel data for birefringence $\Delta m(t)$
(dotted line, representing the smoothed model in Fig.~\ref{Dn}).}
\label{phi-Dn=f(time))}
\end{figure}

We obtain the same results for changes in local order parameter
swelling by a racemic solvent and calculate, from the
Eq.~(\ref{pitch=f(Dn,psi)}), how the effective helical pitch
$\langle \tilde{p}\rangle$ varies with time. The last task
remaining, to successfully map thus obtained $\langle p(t)
\rangle$ on the important concentration dependence, is to
independently measure the change in total concentration $\phi =
\phi(t)$ as the network gradually loses the solvent.
Figure~\ref{phi-Dn=f(time))} shows the result of such weight
measurement of our cholesteric sample initially swollen in a large
amount of each corresponding solvent. Such an experiment is,
necessarily, much less accurate than the rest of our data --
nevertheless the results appear unambiguously pointing at a
different saturation level at $t \rightarrow \infty$.

In fact, the detailed analysis of the decay of total concentration
of solvent in a swollen cholesteric network shows another basic
and interesting result, the discussion of which is beyond the
scope of this paper. We attempted to fit the data for $\phi(t)$
(below $\phi^*$, in the mesophase) and the only good fit was
achieved by a ``stretched exponential'' function, the same for
both solvents (with a relaxation time slightly different, which is
only natural for chemically different solvents):
\begin{eqnarray}
{\rm achiral: } && \phi(t) = C \, e^{-(t/\tau_{\rm a})^{2/3}}
\label{fits} \\
{\rm racemic: } && \phi(t) =  C \, e^{-(t/\tau_{\rm r})^{2/3}} +
\Delta \phi \ , \nonumber
\end{eqnarray}
with $\tau_{\rm a} \approx 2.6$ and $\tau_{\rm r} \approx 1.5$\,h,
but with characteristically the same prefactor $C \approx 14$ (in
\% units, as in Fig.~\ref{phi-Dn=f(time))}). Such a time
dependence of average concentration during surface evaporation is
very different from classical predictions \cite{Crank:75},
offering the law $\phi = (8 \phi_0/\pi^2) \exp [-\pi^2 D t/8d^2]$.
Such a deviation could be due to coupled nonlinear effects of
elastic strain, order $Q$ and inhomogeneous solvent concentration
across the sample layer.

\begin{figure}
\resizebox{0.47\textwidth}{!}{\includegraphics{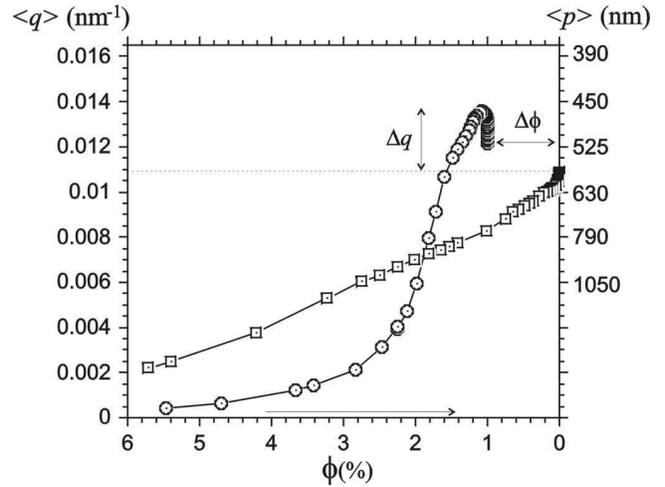}}
\caption{Variation of phase chirality, represented by the helix
wavenumber $\langle q \rangle$ (or, equivalently, the pitch
$\langle \tilde{p}\rangle$ on the right $y$-axis) as function of
decreasing concentration $\phi$, in achiral (data labelled by
squares) and racemic (circles) environments. The increase in
$\Delta q$ and the retention of chiral solvent component $\Delta
\phi$ are indicated on the plot. } \label{p=f(phi))}
\end{figure}

Having mapped the data for total concentration $\phi(t)$ [and the
associated sample thickness decrease, $d_0(1+\phi)$] on the
results for optical rotation, $\Psi(t)$ and local birefringence
$\Delta m(t)$, we can obtain the concentration dependence of phase
chirality in our system. Figure~\ref{p=f(phi))} presents the
results of our analysis, based on the Eq.~(\ref{pitch=f(Dn,psi)}),
for the variation of $\langle \tilde{p} \rangle = 2\pi/\langle q
\rangle$ as function of concentration.\footnote{The results in the
region of phase transition, close to $\phi^*$, are too ambiguous
because of several independent critical functions (especially
$\Delta m$) and not shown in the plot} The data correspond
directly to the results of Fig.~\ref{Dpsi}.

It appears that, unlike for the temperature effect illustrated in
Fig.~\ref{Dn-psi=f(temp)}, a continuous unwinding of the natural
helix occurs on swelling in achiral solvent. This is not an affine
effect of increasing sample thickness: the $d(t)$ variation has
been accounted for in evaluating the rotation rate $d\Psi/dz$. The
likely reason for such an effect is the dilution of molecular
chirality and its reducing effect on the macroscopic scale.

The main effect we are interested in is the stereo-selective
response to the racemic solvent. The marked increase in the
effective phase chirality and the failure of the solvent to
evaporate completely (also demonstrated in
Fig.~\ref{phi-Dn=f(time))}) are clear indications of the network
selecting and retaining the solvent component with the chirality
sense matching that of the helix. We could reasonably assume that
all the retained solvent is left-handed (L), so the $\Delta \phi
(t \rightarrow \infty) = \phi_L \approx 1\%$. Note, however, that
the strength of chiral solvent retention only become noticeable
when the overall solvent content $\phi$ reduces below $\sim 3\%$.
At this level, the local order parameter $Q$ (expressed as $\Delta
m$ in Figs.~\ref{Dn} and \ref{phi-Dn=f(time))}) increases to its
nearly saturation value characteristic of the dry network. When
the local nematic order is weak, at higher $\phi$, we see no
stereo-selectivity. This indicates the role of phase chirality in
the observed phenomenon, as well as the method of extracting the
L-solvent trapped in the network: one simply needs to heat the
material to its isotropic state, or mechanically stretch above the
critical strain so that the helix is unwound \cite{WT03}.

Comparing the final values of $\Delta q$ and $\Delta \phi$, we can
deduce another phenomenological parameter used in the MW theory,
the coefficient $q_{\rm s}$ in the Eq.~(\ref{p}). After the affine
expansion effect is eliminated, as in Fig.~\ref{p=f(phi))}, the
remaining linear relation gives $q_{\rm s} \approx 0.1 \,
\hbox{nm}^{-1}$.

%

\section{Conclusions}
In summary, we found a strong and reproducible effect of
stereo-selectivity in cholesteric elastomers. The study has
unambiguously pointed to the macroscopic phase chirality (the
cholesteric helix) as the dominant force behind the phenomenon.
From a combination of analytical tools provided by the theory of
optical rotation in cholesterics and the local analogy between the
cholesteric and the nematic elastomer systems, we were able to
quantitatively follow the variation of helical pitch with solvent
concentration. Comparing the effects of an ordinary achiral
solvent and the racemic mixture of two opposite chiral
small-molecule components, we demonstrate how the cholesteric
elastomer selectively retains the component with the matching
sense of chirality.

Most of our experimental work was based on delicate optical
measurements and required monodomain cholesteric (and nematic)
elastomers. Preparation and the resulting quality of these could
present many practical difficulties. However, for the purpose of
stereo-selectivity, one does not need monodomain networks! Once we
have demonstrated and studied the effect in model samples, one can
now proceed to develop a new technology of chiral separation using
cholesteric polymer networks of different chemical composition (to
control overall solubility in target racemic solvents and to
choose the desired handedness of the helix). These networks do not
have to be aligned in any way; perhaps the best practical way is
to prepare highly porous sponges with high internal surface area,
which could then be used to extract chiral components from racemic
mixtures (and release them on, e.g., subsequent heating). However,
our results indicate that careful monitoring of timing and overall
solvent content has to be maintained in order for such a ``chiral
sponge'' to remain in its most effective regime (at high local
$Q$). Numerous applications of this effect in biomedical industry
and stereo-selective sensing come to mind.

Several fundamental and interesting problems were left behind. The
most startling is the oscillating effect of solvent loss under
mechanical constraints, but even the basic effects of diffusion in
the medium with coupled non-linear degrees of freedom are
challenging for both theory and experiment. \\

We acknowledge many useful discussions with M. Warner and P.
Cicuta, and the financial support from EPSRC.

\end{document}